\documentclass[12pt]{iopart}
\usepackage{iopams} 
\input epsf
\expandafter\let\csname equation*\endcsname\relax
 \expandafter\let\csname endequation*\endcsname\relax
 \usepackage[fleqn]{amsmath}
\usepackage{amsfonts}
\usepackage{amssymb}
\usepackage{amsopn}
\usepackage{epsfig}
\usepackage{graphics,psfrag,rotating}
\usepackage{graphicx}
\usepackage{dcolumn}
\usepackage{bm}
\usepackage{epstopdf}
\usepackage{color}
\usepackage[usenames,dvipsnames,svgnames]{xcolor}
\usepackage[colorlinks=true,
      linkcolor=red,
      urlcolor=gray,
      citecolor=blue]{hyperref}
 \usepackage{hyperref}
 \usepackage[]{cite}
 \usepackage{subfloat}
\usepackage{subfig}

\def\ga{\gamma}

\def\3nab{\tilde{\nabla}}

\def\hsp5{\hspace{5mm}}

\def\case#1/#2{\textstyle\frac{#1}{#2}}

\def\ber {\begin{eqnarray}}
\def\eer {\end{eqnarray}}
\def\bea {\begin{eqnarray}}
\def\eea {\end{eqnarray}}

\def\bc {\begin{center}}
\def\ec {\end{center}}
\def\case#1/#2{\frac{#1}{#2}}

\newcommand{\bw}{\begin{widetext}}
\newcommand{\ew}{\end{widetext}}
\newcommand{\nn}{\nonumber\\}

\newcommand{\be}{\begin{equation}}
\newcommand{\bse}{\begin{subequation}}
\newcommand{\ese}{\end{subequation}}
\newcommand{\ee}{\end{equation}}
\newcommand{\eei}{\end{eqnarray}\indent\indent}
\newcommand{\ba}{\begin{array}}
\newcommand{\ea}{\end{array}}
\newcommand{\bal}{\begin{eqnarray}}
\newcommand{\eal}{\end{eqnarray}}

\def\case#1/#2{\textstyle\frac{#1}{#2} }

\newcommand{\bver}{\begin{verbatim}}
\newcommand{\ever}{\end{verbatim}}

\begin{document}
\title{A Generalized Solution of Bianchi Type-$V$ Models with Time-dependent $G$ and $\Lambda$}
\author{Alnadhief H. A. Alfedeel$^{1,2}$\footnote{
alfaln001@uofk.edu}, Amare Abebe$^{3}$\footnote{amare.abbebe@gmail.com} and Hussam M. Gubara$^{4}$\footnote{hussam.m.a86@hotmail.com}}
\address{$^1$ Department of Physics, University of Khartoum, P.O. Box 321, Khartoum 11115, Sudan}
\address{$^{2}$ Information Technology Department, Napata College, Ryad, Khartoum, Sudan}
\address{$^{3}$ Center for Space Research \& Department of Physics, North-West University, Mafikeng 2735, South Africa}
\address{$^{4}$ Faculty of Mathematical Sciences \& Statistics, Al-Neelain University, Khartoum, Sudan}
\date{\today}

\begin{abstract}
We study the homogeneous but anisotropic Bianchi type-$V$ cosmological model
with time-dependent gravitational and cosmological ``constants''. Exact solutions
of the Einstein field equations (EFEs) are presented in terms of adjustable
parameters of quantum field theory in a spatially curved and expanding background.
It has been found that the general solution of the average scale factor $a$
as a function of time involved the hypergeometric function. Two cosmological models
are obtained from the general solution of the
hypergeometric function and the Emden--Fowler equation. The analysis of the models shows that, for a particular choice of parameters in our first model,
the cosmological ``constant'' decreases whereas the Newtonian gravitational ``constant'' increases with time, and for another choice of parameters, the opposite behaviour is observed.   The models become isotropic at late times for all parameter choices of the first model. In the second model of the general solution, both the cosmological and gravitational ``constants'' decrease while the model becomes more anisotropic over time.
The exact dynamical and kinematical quantities have been calculated analytically for each model.
\end{abstract}

 
 \maketitle


\section{Introduction}
The Bianchi models are described as classes of non-standard cosmological models that are in principle spatially homogeneous but anisotropic. Alternatively, they can be considered as a generalization of the well-known standard Friedman--Lema\^{i}tre--Robertson--Walker (FLRW) models of cosmology. Originally,  these models date back to the work of Bianchi \cite{Bianchi1, Bianchi2}, who classified them according to their construction of homogeneous surfaces in space-time. These surfaces are constructed by the action of a three-dimensional group of isometrics $G_3$ upon the space-like $3-$surfaces \cite{Ellis2006}. From the cosmological point of view, these models are of great interest because they provide a way of studying the anisotropy at an early period of our universe's expansion history \cite{Coles2002}.

One of the most puzzling and unsolved problems in physics today is the Cosmological Constant Problem \cite{Weinberg1989} in the Einstein field equations (EFEs). In cosmology, it is regarded as a matter field with negative pressure (or as a vacuum energy density) that drives the accelerated expansion of the universe. The calculations of quantum field theory predict the value of vacuum to be of the order of $10^{130}$ s$^{-2}$, whereas the cosmological constant value is of the order of $10^{-20}$~s$^{-2}$; hence, a new thought is required to explain or resolve this puzzle. The idea of considering the cosmological models as varying vacuum energy density $\Lambda$ has extensively been adapted by several authors \cite{Vishwakarma1996a, Vishwakarma1996b, Vishwakarma1999, Vishwakarma2000, Vishwakarma2001,Vishwakarma2005, Berman1990a, Berman1989, Berman1990b, Berman1991a, Berman1991b, Arbab1997, MishraRavi2014, AruneshMishra2013, Anirudh2013, Chawla2012,PradhanHassan2011}. Chen~\cite{Chen} considered $\Lambda$ proportional to $1/a^{2}$, and a generalized form
$\Lambda = \alpha/ a^{2} + \beta H^2 $, which depends on adjustable parameters $\alpha$ and $\beta $ of the quantum field on a curved and expanding background, the Hubble parameter $H$ and the average scale factor of the universe $a$ was also introduced \cite{Carvalho}.

Since the Newtonian constant $G$ is a coupling ``constant'' between the geometry of spacetime and energy in the general theory of relativity (GR), and the universe is evolving with time, it is natural to assume that G varies with time. This consideration was first introduced by Dirac \cite{Dirac1937}. Based on this idea, there have been many attempts to modify the theory of GR, but, unfortunately,
one of these efforts  has yet to be universally accepted or studied extensively. Recently, there has been an increasing interest in studying modifications of GR with variable cosmological and Newtonian ``constants'' \cite{Abdel-Rahman1990,Pradhan2005, Kalligas1992, Abdussattar1997,Singh2007,Borges2005,Singh2008,Singh}. Apart from these investigations that include the variation in both $G$ and $\Lambda$ within the limit of GR, there are also studies  on cosmological models with viscous fluids in the presence of $G$ and $\Lambda$  \cite{Arbab1997, Arbab1998, Prashant2012}.

Many authors have studied the solutions of EFEs for homogeneous and anisotropic Bianchi type models \cite{Pradhan2001,Pradhan2004, Pradhan2005, Pradhan2006, Hajj1985,Hajj1986, Shri Ram1989, Shri Ram1990,Camci, Mazumder1994, Tiwari2008, Tiwari2009}. For instance, Dwivedi \cite{Dwivedi2012} has solved the EFEs for a Bianchi Type-$V$ model with variable cosmological and Newtonian ``constants“ for a stiff perfect fluid. In his solution, the physical and  kinematical parameters have been fully described, and he has found that the model has a singularity point, and all the physical parameters decrease as
time increases, with the model isotropizing at late times. In addition to that, the universe described by such a model expands at a constant rate (i.e., the deceleration parameter equals zero). On the other hand, Yadav \cite{Yadav:2009cj} has solved the same problem and generated two types of solution for the average scale factor, namely power-law type and exponential-type solutions. He found that the ``cosmological constant'' decreases with time and it reaches a small positive value at late and early times, respectively, results supported by supernova Type Ia observations.  It is now well established from a variety of studies \cite{Dwivedi2012} that all the previous solutions are for special cases where the values of $\alpha$ and $\beta$ are chosen from the beginning of the solution process, and no previous study has investigated the general case.

The main aim of this paper is to find the general solution of the EFEs for Bianchi type-$V$ models for a stiff perfect fluid with variable $\Lambda$ and $G$ without making any constraints on the value of $\alpha$ and $\beta$ in the $\Lambda$ term, and to describe the behavior of the physical and kinematical parameters of the models. The remaining part of the paper is structured as follows: Section \ref{bianchiv} is dedicated for a short description of the Bianchi type-$V$ model and their general mathematical solutions for the EFEs.
In Section \ref{models}, we present our models for restricted values of  either $\alpha$ or $\beta$, and considering the other parameter as free. Finally, we conclude our results in Section~\ref{conc}.

We follow the Misner--Wheeler--Thorne notation where $x^a= (-x^0,x^1,x^2,x^3)$, Latin letters for the metric and geometric units with $c=1$. 

\section{Bianchi Type-\emph{V} Cosmology}\label{bianchiv}
We consider the spatially homogeneous and anisotropic Bianchi type-$V$ space-time that is represented by
the following line-element:
\begin{equation}\label{metric}
ds^2 = -dt^2 + A^2(t) dx^2 + e^{2 x} \left[ B^2(t) dy^2 + C^2(t) dz^2 \right],
\end{equation}
where $A(t), B(t)$ and $C(t)$ are the components of the fundamental metric tensor. We assume that the
cosmic matter is a perfect fluid that is represented by the following energy-momentum tensor:
\begin{equation}
T_{ij} = ( p + \rho ) u_{i} u_{j} + p g_{ij}~,\label{EnergyTesor}
\end{equation}
where $\rho$ is matter density, $u^i = \delta^i_t = (-1,0,0,0)$ is the normalized fluid four-velocity, which
is a time-like quantity such that $u^iu_i=-1$, and $p$ is the fluid's isotropic pressure. $\rho$ and $p$ are related through the barotropic
equation of state
\begin{equation}\label{Eqstate}
p= w \rho~, \qquad 0\leq w \leq 1\;,
\end{equation}
where $w$ is the equation-of-state (EoS) parameter.
The EFEs with time-dependent $\Lambda$ and $G$ are given by
\begin{equation}\label{EFEs}
R_{ij}-\frac{1}{2} g_{ij} R = -8 \pi G(t) T_{ij} + g_{ij}\Lambda(t)\;,
\end{equation}
where $R_{ij}$ and $g_{ij}$ are the Ricci and metric tensors, respectively, and $R$ is the Ricci scalar.
Substituting Equations \eqref{metric} and \eqref{EnergyTesor} into Equation \eqref{EFEs}, we get the field equations
\begin{align}
\frac{ \dot{A} \dot{B} }{AB} + \frac{ \dot{B} \dot{C} }{BC} + \frac{ \dot{A} \dot{C} }{AC} - \frac{3}{A^{2}}
& = 8 \pi G(t) \rho + \Lambda(t)~, \label{Gtt}\\
\frac{ \ddot{B} }{B} + \frac{ \ddot{C} }{C} + \frac{ \dot{B} \dot{C} }{BC} - \frac{1}{A^{2}} 
& = -8\pi G(t) p + \Lambda (t)~,\label{Gxx} \\
\frac{ \ddot{C} }{C} + \frac{ \ddot{A} }{A} + \frac{ \dot{C} \dot{A} }{CA} - \frac{1}{A^{2}} 
& =   -8\pi G(t) p +\Lambda (t)~, \label{Gyy} \\
\frac{ \ddot{A} }{A} + \frac{ \ddot{B} }{B} + \frac{ \dot{A} \dot{B} }{AB} - \frac{1}{A^{2}} 
& =  -8\pi G(t) p +\Lambda  (t)~,\label{Gzz}\\
\frac{2\dot{A}}{A} - \frac{\dot{B}}{B} - \frac{\dot{C}}{C} & = 0~,\label{Gxt}
\end{align}
with the dot over a letter representing differentiation with respect to time. The covariant divegence of the left hand side of 
Equation \eqref{EFEs} produces
\begin{equation}\label{G+rho+Evolution}
8\pi G\left[\dot{\rho} + (\rho+ {p})\left(\frac{\dot{A}}{A} 
+\frac{\dot{B}}{B}+\frac{\dot{C}}{C}\right)\right]+8\pi\dot{G}+\dot{\Lambda}=0~,
\end{equation}
while the conservation of the usual energy-momentum tensor $T^{ij}$ (i.e., $\nabla_{j} T^{ij}=0$) yields  
\begin{equation}\label{EvolEq}
\dot{\rho}+(\rho+ {p})\left(\frac{\dot{A}}{A}+\frac{\dot{B}}{B}+\frac{\dot{C}}{C}\right)=0\;.
\end{equation}
Substituting Equation \eqref{EvolEq} into \eqref{G+rho+Evolution}, we get   
\begin{equation}\label{GEvoEq}
8\pi\rho\dot{G}+\dot{\Lambda}=0.\;
\end{equation}
This equation shows how $\Lambda$ and $G$ evolve with time. It also shows that the two ``constants'' form a coupled system and, therefore, do not evolve independently of each other.
The average scale factor $a=a(t)$ for Bianchi-$V$ models is defined to be    
\begin{equation}\label{AverageScalefactor}
 a =(ABC)^{1/3}\;,
\end{equation}
and the generalized Hubble parameter $H$ is defined as in \cite{Dwivedi2012} 
\begin{equation}\label{generalizedH}
H=\frac{\dot{a}}{a}= \frac{1}{3} \left( \frac{\dot{A}}{A}+ \frac{\dot{B}}{B}+\frac{\dot{C}}{C} \right)
= \frac{1}{3}(H_1+H_2+H_3)\;,
\end{equation}
where $H_1, H_2$ and $H_3$ are directional Hubble's parameters along $x$, $y$ and $z$ directions, respectively. The deceleration parameter $q$ follows the usual definition:
\begin{equation}\label{generlizedq}
q=- \frac{\ddot{a} a}{\dot{a}^2 }=-1-\frac{\dot{H}}{H^2}.\;
\end{equation}

The volume expansion parameter $\theta$, the average anisotropy parameter $A_p$, and
shear modules $\sigma$ are defined as in \cite{Dwivedi2012}
\ber
&&\theta= \nabla_{i} u^i = 3 H~, \label{volumeexpansion}\\
&&A_p  = \frac{1}{3}\sum_{i=1}^{3} \left(\frac{H_i - H}{H} \right)^2~,\\
&&\sigma^2  = \frac{1}{2} \sigma_{ij}\sigma^{ij}  = \frac{1}{2} \left( \frac{\dot{A}^2} {A^2} + \frac{\dot{B}^2 }{B^2} + \frac{\dot{C}^2 }{C^2} \right) - \frac{ \theta^2}{6}~ \nn \\ 
 &&~~~~ = \frac{1}{3} \left( \frac{\dot{A}^2} {A^2} + \frac{\dot{B}^2 }{B^2} + \frac{\dot{C}^2 }{C^2}\right)-\frac{1}{3}\left(\frac{\dot{A} \dot{B} } {A B} + \frac{\dot{B} \dot{C}}{B C} + \frac{\dot{A} \dot{C} }{A C}  \right)~,
    \label{shear}
\eer
where the term $\sigma^{ij}$ represents the shear tensor.  For this model, its scalar quantity comes out to be
\begin{equation}
\sigma=\frac{K}{a^3}\;,\label{shearK}
\end{equation}
where $K$ is a positive constant that is related to the anisotropy of the model. Having introduced these quantities, we can re-express the field Equations
\eqref{Gtt}--\eqref{Gzz} and \eqref{EvolEq} in terms of $a$, $H$, $q$ and $\sigma$ as
\ber\label{GxxGyyGzz}
&&8\pi G {p}-\Lambda=(2q-1)H^2-\sigma^2+\frac{1}{A^2}\;,\\
&&\label{Gttshq} 8\pi G\rho+\Lambda =3H^2-\sigma^2-\frac{3}{A^2},
\eer
so that, when subtracting Equation \eqref{GxxGyyGzz} from Equation \eqref{Gttshq}, we eliminate the $\sigma$ term, obtaining
\begin{equation}
\frac{\ddot{a}}{a} + 2 \frac{\dot{a}^2}{a^2} - \frac{2}{A^2} = 4 \pi G(t)\rho(1-w) + \Lambda(t)~.\label{RdiffEq}
\end{equation}

We see that this equation cannot be integrated as it is because of the unknown functions  
$A$, $G$, $\rho$ and the parameter $w$. Now, integrating Equation \eqref{Gxt}
and absorbing the integration constant into $A$ or $B$, gives  
\begin{equation}
 A^2 = BC\;,
\end{equation}
and substituting it into Equation \eqref{AverageScalefactor}, we see that 
the average scale factor reads  
\begin{equation}
 a=A~. \label{AEq}
\end{equation}
Subtracting Equation \eqref{Gxx} from \eqref{Gyy} gives
\begin{align}
\frac{ \ddot{B} }{B}  +  \frac{ \dot{B} \dot{C} }{BC} = \frac{ \ddot{A} }{A} + \frac{ \dot{C} \dot{A} }{CA} \longrightarrow
\frac{ C \ddot{B}   + \dot{B} \dot{C} }{BC}  = \frac{ C \ddot{A} + \dot{C} \dot{A} }{CA} ~,
\end{align}
which is equivalent to
\begin{align}
\frac{1}{BC} \frac{d}{dt}\left( C \dot{B}\right) & = 
\frac{1}{AC} \frac{d}{dt}\left( C \dot{A} \right) ~, 
\end{align}
and, multiplying this equation by $ABC$, we obtain 
\begin{align}
A\; \frac{d}{dt}\left( C \dot{B}\right)  & = 
B\; \frac{d}{dt}\left( C \dot{A} \right)\;.
\end{align}
Both sides of this equation are a factorization of a product rule of the following  
\begin{align}
\frac{d \left( A \; (C \dot{B})\right) }{dt} - \dot{A}\;\dot{B} \;C  & = 
 \frac{d \left(B\;( C \dot{A}) \right) }{dt}-  \dot{A}\;\dot{B} \;C\;.
\end{align}
Eliminating $\dot{A}\;\dot{B} \;C$ from both sides and integrating with respect to $t$ yields
\begin{align}
 A \; (C \dot{B}) & = 
B\;( C \dot{A}) + k_1~ , 
 \end{align}
and dividing throughout by $ABC$ yields
\begin{align}
 \frac{ \dot{B}}{B} & = 
\frac{\dot{A}}{A} + \frac{k_1}{ABC}~ , 
 \end{align}
where $k_1$ is a constant of integration. Recalling that $A=a$, we finally get
a first-order coupled differential equation of $A$ and $B$ as follows:
 \begin{align}
\frac{\dot{B} }{B} & = \frac{ \dot{A} }{A} + \frac{ k_1 }{ a^3 }\;.\label{BAdot}  
\end{align}

Similarly, the subtraction of \eqref{Gxx} from \eqref{Gzz}, and \eqref{Gyy}
from \eqref{Gzz}, respectively, gives
\begin{align}
\frac{\dot{C}}{C} & = \frac{\dot{B}}{B} + \frac{k_3}{a^3}~,\label{CBdot} \\
\frac{\dot{C}}{C} & = \frac{\dot{A}}{A} + \frac{k_2}{a^3}~,\label{CAdot} 
\end{align}
where $k_2$ and $k_3$ are constants of integration. Again
integrating the above set of Equations \eqref{BAdot}--\eqref{CBdot}, we get
\begin{align}
 \frac{B}{A} &= d_1 \exp\left(k_1\int\frac{dt}{a^3}\right)~,\label{eq318}\\
 \frac{C}{B} &= d_2 \exp\left(k_2\int\frac{dt}{a^3}\right)~,\label{eq318}\\
 \frac{C}{A} &= d_3 \exp\left(k_3\int\frac{dt}{a^3}\right)\;.\label{eq319}
\end{align}

These equations can be combined to give 
\begin{align}
B &= m_1\;a\; \exp \left(k \int\frac{dt}{a^3}\right)~,\label{Beq}\\
C &= m_2\;a\; \exp\left(-k \int\frac{dt}{a^3}\right)\;,\label{Ceq}
\end{align}
where $m_1\;,m_2$ and $k$ are constant values that depend on
the undetermined constants $d_1,d_2$ and $d_3$, see~\cite{CPSingh2009} for more details. It is clear that Equations \eqref{AEq}, \eqref{Beq} and \eqref{Ceq} are totally dependent on the value of $a$, which is yet to be determined. 
To complete the solution process, we substitute Equation \eqref{AEq} into 
Equation \eqref{RdiffEq} so that  
\begin{equation}
\frac{\ddot{a}}{a} + 2 \frac{\dot{a}^2}{a^2} - \frac{2}{a^2} = 4 \pi G(t)\rho(1-w) + \Lambda(t)~.\label{RdiffEq1}
\end{equation}

Now, following \cite{Carvalho} for the parametrization of $\Lambda$, we assume 
\begin{align}
\Lambda(t) = \frac{\alpha}{a^2} + \beta H^2\;.\label{Lambda}
\end{align}

Setting $w=1$ for a stiff cosmological fluid and substituting Equation \eqref{Lambda} into Equation \eqref{RdiffEq1} produces 
\begin{equation}
\frac{ \ddot{a} }{ a } + (2-\beta) \frac{ \dot{a}^2 }{ a^2 }   - \frac{1}{ a^2 } (2 + \alpha) =0 \;,\label{RdiffEq2}
\end{equation}
which can also be re-arranged as
\begin{equation}
a \ddot{a}  + (2-\beta) \dot{a}^2 - (\alpha  +2) = 0\;.\label{RDE}
\end{equation}

This is the generalized Friedman equation of Bianchi type-$V$ models. The first term $a \ddot{a}$ represents force per unit mass times distance or work of a system that is equivalent to a potential energy, the~second term $(2-\beta) \dot{a}^2$ is a kinetic energy and $(2-\beta)$
is a kind of mass, and the last term is the total energy of the system.    

\begin{figure}[h!]
  \centering
  \subfloat[the evolution of $\Lambda(t)$ in time.]{\includegraphics[width=0.45\textwidth]{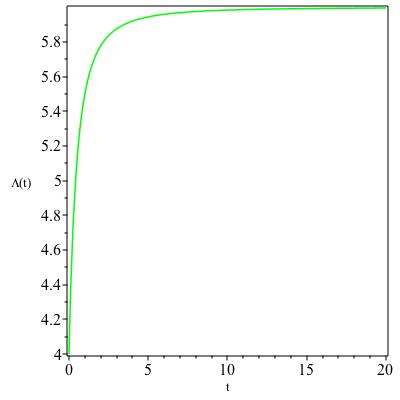}\label{figl0}}
  \subfloat[the evolution of $G(t)$ in time.]{\includegraphics[width=0.45\textwidth]{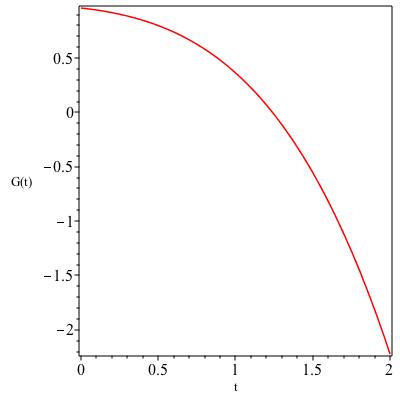}\label{figg0}}
  \caption{The variation of $\Lambda$ and $G$ for $\alpha=-2\;$, $\beta=2\;.$ In (a), we see that an initially increasing $\Lambda$ asymptotically reaches a constant value at late times, whereas (b) shows an initially positive $G$ decreases with time, attaining negative values at alate times.}
      \label{fig1}
\end{figure}

\begin{figure}[h!]
  \centering
  \subfloat[the evolution of $V(t)$ in time.]{\includegraphics[width=0.45\textwidth]{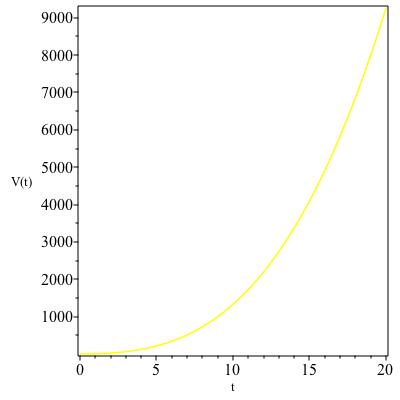}\label{figv0}}
  \subfloat[the evolution of $\sigma(t)$ in time.]{\includegraphics[width=0.45\textwidth]{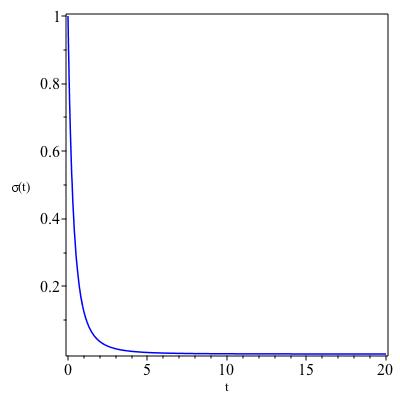}\label{figs0}}
    \caption{The variation of $V$ and $\sigma$ for $\alpha=-2\;$, $\beta=2\;.$ In (a), we see a monotonically increasing volume due to the expanding scale factor, whereas (b) shows an initially anisotropic universe rapidly isotropizing at alate times.}
      \label{fig2}
\end{figure}

As we have mentioned that $\alpha$ and $\beta$ are adjustable constants of quantum field theory in an expanding curved background, the challenging problem here is to solve Equation \eqref{RDE} for general values of $\alpha$ and $\beta$ without making any constraints from the beginning as some authors did \cite{Dwivedi2012, Yadav:2009cj}. It is obvious that, for $\alpha\neq -2,\beta\neq 2$, Equation \eqref{RDE} is a nonlinear second-order differential equation, which 
is not easy to solve unless we transform it into a set of first-order equations. 
Before we move to the general solution, however, we remark that, for the special cases of $\alpha=-2,\beta= 2$, the model has a scale factor solution that grows linearly with cosmic time, i.e.,
\be
a(t)=C_1t+a_0\;,
\ee
where $C_1$ and $a_0$ are integration constants. This solution has a constant expansion and an initially increasing $\Lambda$ that asymptotes to a constant value at late times, as well as a decreasing $G$, as can be seen in Figure \ref{fig1}. Moreover, we can see from Figure \ref{fig2} that the model describes an expanding universe with an overall increasing volume and asymptotically approaches isotropy at late times.
\unskip

\begin{figure}[h!]
  \centering
  \subfloat[the evolution of $\Lambda(t)$ in time.]{\includegraphics[width=0.45\textwidth]{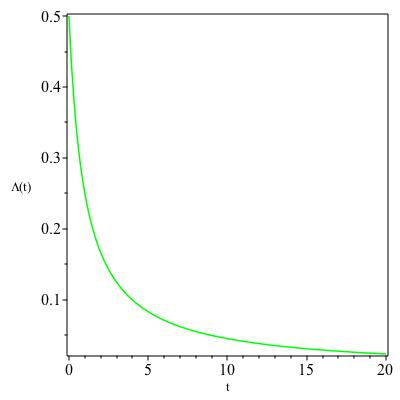}\label{figl01}}
  \subfloat[the evolution of $G(t)$ in time.]{\includegraphics[width=0.45\textwidth]{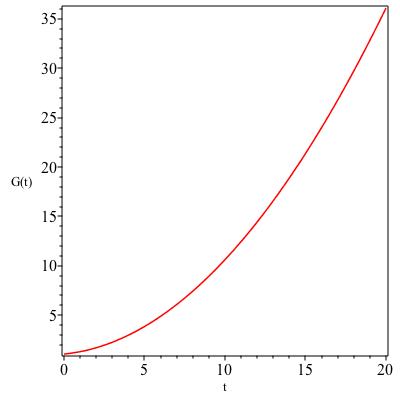}\label{figg01}}
  \caption{The variation of $\Lambda$ and $G$ for $\alpha=-2\;$, $\beta=1.$ For this model, we see a monotonically decreasing $\Lambda$ in Panel (a), whereas Panel (b) shows an increasing $G$ with time.}
      \label{fig3}
\end{figure}

\begin{figure}[h!]
  \centering
  \subfloat[the evolution of $\Lambda(t)$ in time.]{\includegraphics[width=0.45\textwidth]{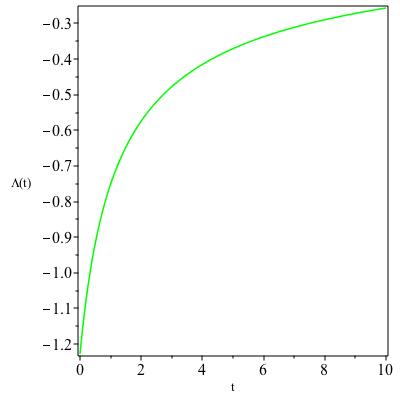}\label{figl01-2}}
  \subfloat[the evolution of $G(t)$ in time.]{\includegraphics[width=0.45\textwidth]{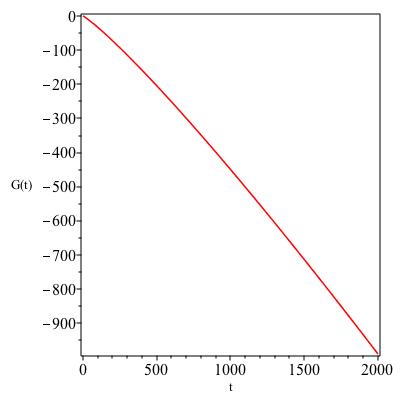}\label{figg01-2}}
  \caption{The variation of $\Lambda$ and $G$ for $\alpha=-2\;$, $\beta=-0.5.$. Here the opposite behaviour to Figure \ref{fig3} is observed, namely an increasing $\Lambda$ and a decreasing $G$ for the choice of model parameters made.}
      \label{fig3-2}
\end{figure}

\begin{figure}[h!]
  \centering
  \subfloat[the evolution of $V(t)$ in time.]{\includegraphics[width=0.5\textwidth]{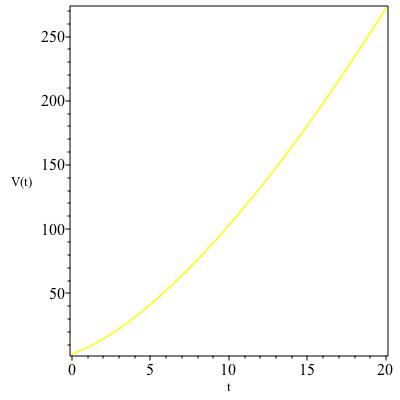}\label{figv01}}
  \subfloat[the evolution of $\sigma(t)$ in time.]{\includegraphics[width=0.5\textwidth]{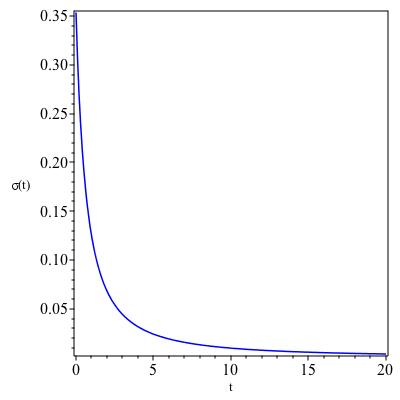}\label{figs01}}
  \caption{The variation of $V$ and $\sigma$ for $\alpha=-2\;$, $\beta=1\;.$ The left panel depicts a monotonically increasing volume of the the universe (as would be expected due to the cosmological expansion) whereas the right panel depicts a universe isotropizing at late times.}
      \label{fig4}
\end{figure}

On the other hand, for $\beta\neq 2\;,3$ and $\alpha=-2$, the model provides a solution of the form
\be
a(t)=\left[(3-\beta)\left(C_2t+C_3\right)\right]^{\frac{1}{3-\beta}}
\ee
for integration constants $C_2$ and $C_3$. Unsurprisingly, this solution reduces to the linear expansion solution above when we fix $\beta=2$. We clearly notice that  the behaviours of $\Lambda$ and $G$ depend on the choice of $\beta$. For example, while fixing $\beta=1$ results in Figure \ref{fig3}, a different behaviour is observed when $\beta=-0.5$, as shown in Figure \ref{fig3-2} where $\Lambda$ asymptotically attains a constant value.  It is worthwhile noticing, however, that the general behaviour of $V$ and $\sigma$ remains the same as shown in Figure \ref{fig4} (infinitely increasing and asymptotically vanishing at time infinity, respectively).
\unskip
\unskip

Now, let us introduce an intermediate variable $P$ such that
\be
P = \dot{a}\implies\ddot{a} = \frac{d P}{d a} \; \frac{d a}{d t} = P \;\frac{d P}{d a}\;,\label{RPEq}
\ee
and use this in Equation \eqref{RDE} to obtain
\begin{equation}
 a \; P\;\frac{d P}{d a} + (2-\beta) \; P^2 - (\alpha  + 2 ) = 0\;.\label{RDE0}
\end{equation}

Separating the variables, we can rewrite this as  
\begin{equation}
\frac{P\; d P}{ (\beta-2) \; P^2  + (\alpha  + 2 ) }  =  \frac{da}{a}   ~, \label{RDE1}
\end{equation}
and integrating both sides gives 
\begin{align}
 \ln \left[  \frac{(\beta-2)}{\ga} \; P^2  + \frac{(\alpha  + 2 )}{\ga} \right ]  &=  \ln \left[ a^{2 (\beta-2) }\right]\;,
\end{align}
where $\ga$ is an integration constant. Substituting the value  of $P$ from \eqref{RPEq} and taking the anti-logarithm for both sides gives  
\begin{align}
    \dot{a}^2 - \frac{\ga}{(\beta-2)} \; a^{2 (\beta-2)} = \left( \frac{\alpha  + 2}{2-\beta} \right)\;.\label{R2dot} 
\end{align}

Integrating Equation \eqref{R2dot}, we can see that
\begin{equation}
 \int\frac{ d a } { \sqrt{ \frac{\ga}{(\beta -2 )} \; a^{ 2(\beta-2)} + \left( \frac{\alpha  + 2}{2-\beta} \right) } } =
 \frac{a}{ \sqrt{ \frac{\alpha+2}{2- \beta} } }
\; _2 F_1 \left( \frac{1}{2}, \frac{1}{2(\beta-2)}, 1 + \frac{1}{2(\beta-2)}; \frac{\ga}{(\alpha+2)} a^{2(\beta-2)} \right)
= \pm\tau\;, \label{RtHyperfunc}
\end{equation}
which can be rewritten as 
\begin{equation}
a\; _2 F_1 \left( \frac{1}{2}, \frac{1}{2(\beta-2)}, 1 + \frac{1}{2(\beta-2)}; \frac{\ga}{(\alpha+2)} a^{2(\beta-2)} \right)
= \pm\sqrt{ \frac{\alpha+2}{2- \beta}} \tau~, \label{RtHyperfunc2}
\end{equation}
where $\beta\neq2$ and $\alpha \neq-2$, $\tau=t-t_0$ and $_2 F_1$ is the hypergeometric function. 
It is obvious that Equation~\eqref{RtHyperfunc2} is a general form of the solution, 
and cannot be simplified into an analytical expression for $a$ as an explicit function of time
unless we have to make some gauge choices on the values of $\alpha$ and~$\beta$.

Our approach is different from the other previous works of some authors, where they
firstly fixed the value of $\alpha$ and $\beta$ in the $\Lambda$ term, then solved the generalized Friedmann equation, but
we do not assume a priori the value or functional  dependence
of these constants or any other parameters in our~models.
\section{Some Specific Models of the General Solution}\label{models}
In this section, we will present our cosmological models that emerge from the choice of  suitable values of $\alpha$ and $\beta$.
\subsection{Model I: $\beta=1$}
In this model, we choose $\beta=1$  such that Equation \eqref{RtHyperfunc2} gives a power solution of $a$ as
\begin{align}
a = \sqrt{\alpha+2} \left[ \tau^2 +  \frac{\ga}{(\alpha+2)^2} \right]^{\frac{1}{2}}\;.\label{model2}
\end{align}

\begin{figure}[h!]
  \centering
  \subfloat[the evolution of $H(t)$ in time.]{\includegraphics[width=0.45\textwidth]{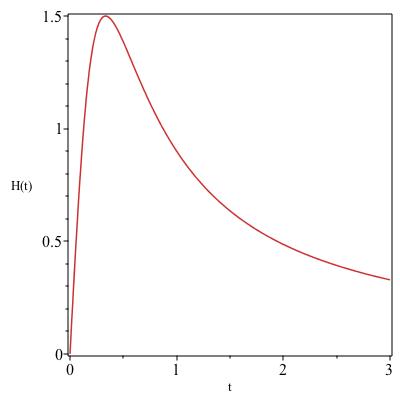}\label{figh1}}
  \subfloat[the evolution of $q(t)$ in time.]{\includegraphics[width=0.45\textwidth]{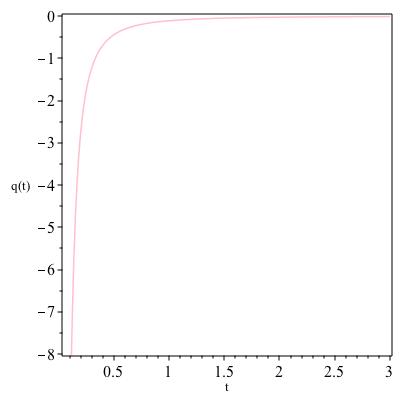}\label{figq1}}
  \caption{The variation of $H$ and $q$ for Model I. For the model with $\alpha\;, \beta$ and $\ga$ all normalized to unity,  we get an initially accelerated expanding solution reaching a finite maximum value and eventually contracting asymptotically towards a vanishing $q$ as depicted by the left and right panels above.}
      \label{fig5}
\end{figure}

\begin{figure}[h!]
\centering
\includegraphics[scale=0.8]{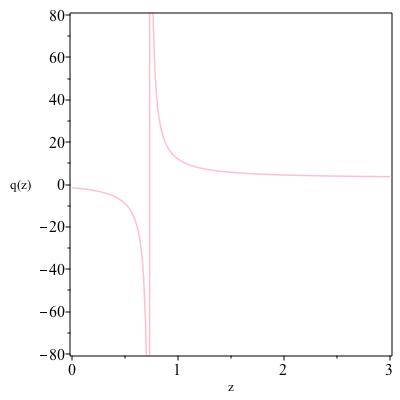}
\caption{The deceleration parameter in terms of redshift for $\alpha=1=\ga$.}
\label{figredshift}
\end{figure}

\begin{figure}[h!]
  \centering
  \subfloat[the evolution of $\Lambda(t)$ in time.]{\includegraphics[width=0.45\textwidth]{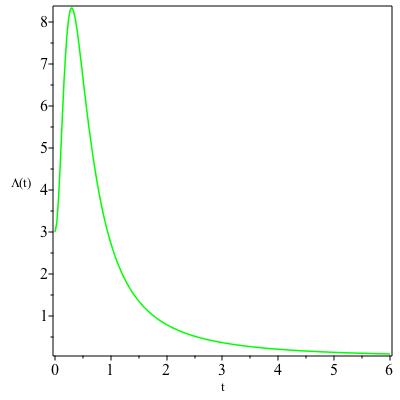}\label{figl1}}
  \subfloat[the evolution of $G(t)$ in time.]{\includegraphics[width=0.45\textwidth]{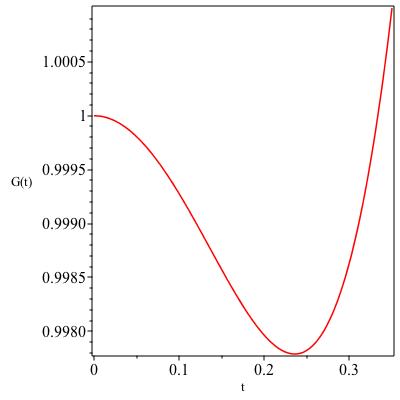}\label{figg1}}
  \caption{The variation of $\Lambda$ and $G$ for Model I.  Panel (a) shows an initially positive $\Lambda$  increasing towards a finite maxumumm value and eventually vanishingly decreasing. Panel (b) shows figure drawn so as to zoom in the decreasing behaviour of G for small t, increasing for large t.}
      \label{fig6}
\end{figure}

\begin{figure}[h!]
  \centering
   \subfloat[the evolution of $V(t)$ in time.]{\includegraphics[width=0.45\textwidth]{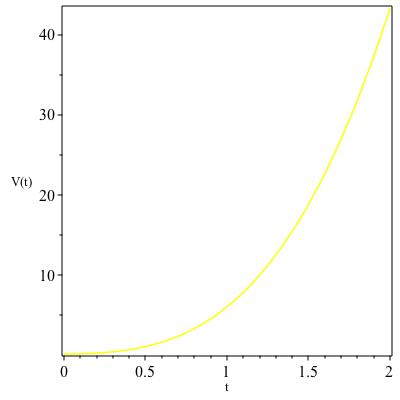}\label{figv1}}
  \subfloat[the evolution of $\sigma(t)$ in time.]{\includegraphics[width=0.45\textwidth]{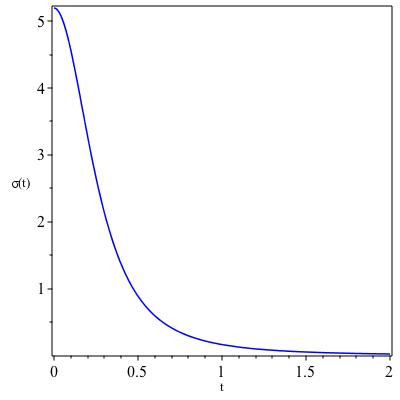}\label{figs1}}
  \caption{The variation of $V$ and $\sigma$ for Model I. Here as in Figures \ref{fig2} and \ref{fig4}, albeit with different slopes, the left panel depicts a monotonically increasing volume of the the universe whereas the right panel depicts a universe isotropizing at late times.}
      \label{fig7}
\end{figure}

Having known the value of $a$ in this model, then the kinematical and dynamical parameters $H\;,\theta\;,\sigma\;,q$, and $\rho$, the Newtonian gravitational ``constant" $G$, and the metric variables $A\;,B\;,C$ are calculated as 

\begin{align}
 H & = \frac{\dot{a}}{a} = \frac{\tau}{\tau^2 +  \dfrac{\ga}{(\alpha+2)^2}}~,\\
\theta & = 3 H = \frac{3\;\tau}{\tau^2 +  \dfrac{\ga}{(\alpha+2)^2}}~,\\
\sigma & = \frac{K}{a^3}= \frac{K}{\left[ \sqrt{\alpha+2} \sqrt{ \tau^2 +  \dfrac{\ga}{(\alpha+2)^2} } \right]^{3}}  ~, \\
\rho &= \frac{\rho_0}{a^6} = \frac{\rho_0}{\left[ (\alpha+2)\tau^2+ \dfrac{\ga}{\alpha+2} \right]^{3}} 
          ~, \\
G & = - \int \frac{\dot{\Lambda}}{8\pi\; \rho} dt=  \frac{{\alpha}^{3} + 5{\alpha}^{2} + 8\alpha + 4}{8\pi\rho_0} \; \tau^{4}
     - \frac{\ga}{4\pi\rho_0} \; \tau^{2} + G_0~,\\
q & =  -\frac{\ddot{a} a}{\dot{a}^2} =-\frac{\ga}{(\alpha+2)^2\tau^2}\;,\\
A & =  \sqrt{\alpha+2} \sqrt{ \tau^2 +  \dfrac{\ga}{(\alpha+2)^2} }~,\\
B &= B_0 \sqrt{\alpha+2} \sqrt{ \tau^2 +  \dfrac{\ga}{(\alpha+2)^2} }  
\; \exp \left[ \frac{k}{ \sqrt{\alpha+2} }\; \frac{ \tau }{ \sqrt{ \tau^2 +  \dfrac{\ga}{(\alpha+2)^2} }} \right] 
~,\\    
C &= C_0 \sqrt{\alpha+2} \sqrt{ \tau^2 +  \dfrac{\ga}{(\alpha+2)^2} }  
\; \exp \left[- \frac{k}{ \sqrt{\alpha+2} }\; \frac{ \tau }{ \sqrt{ \tau^2 +  \dfrac{\ga}{(\alpha+2)^2} }} \right]\;,
\end{align}
where $B_0=m_1e^{\mbox{const}}\;,C_0=m_2e^{\mbox{const}}$ and $G_0$ are constants of integration. 
This model has an initial point of singularity at $t=t_0$ and $\ga=0$. Furthermore, although early-time behaviour, shown in Figures  \ref{fig5}--\ref{fig7}, differs slightly for different choices of $\alpha$ and $\ga$, it can be shown that, as $t\to\infty$,  
$H,\theta, \sigma, \Lambda$, the magnitude of $q$ and $\rho$ all decrease with time, whereas  $G$, the volume $V$ and and the metric variables $A\;,B$ and $C$ are increasing functions of time. It is worth mentioning here that although the numerical values of $\Lambda$ and $q$ decrease asymptotically towards constant values, whether the model describes an accelerated or decelerated expansion solely depends on the sign of $\ga$. For example, a positive choice of $\ga$ describes an early accelerated expansion that eventually slows down to an asymptotically constant expansion expansion, whereas a negative $\ga$ describes an early decelerated expansion that eventually asymptotes to a constant expansion at late times.  In addition, $|\sigma/\theta| \to 0$ as $t\to \infty$, thus indicating that the model approaches isotropy for large values of $t$, as observed in \cite{Dwivedi2012} as~well.

In redshift space, we can show that the deceleration parameter for the model can be given by
\be
q(z)=\frac{\ga(\alpha+2)(1+z)^2}{\ga(1+z)^2-(\alpha+2)}\;,
\ee
where $1+z\equiv \frac{a_0}{a}$.
In Figure \ref{figredshift}, we show in redshift space the transition from an early deceleration epoch (at large redshifts) to late-time acceleration (at small redshifts).

\unskip

\unskip
Here, a mathematical singularity occurs at  the value of $z$ for which $\ga(1+z)^2-(\alpha+2)=0$, and hence the singular point can be shifted either way by choosing appropriate values of $\alpha$ and $\ga$.
\unskip

\subsection{Model II: The Emden--Fowler Approach}
This model naturally arises when comparing the nonlinear Equation \eqref{R2dot} 
to the first integral of the Emden--Fowler equation \cite{Emden}
\be
\ddot{a}=rt^n\;a^m\;,
\ee 
which, for $n=0$, is given by
\begin{align}
   \dot{a}^2- \frac{2 r}{m+1}\;a^{m+1} = s\;,\label{SpcaseEmden}
\end{align}
where $m\;,n\;,r$ and $s$ are constants. This equation has a particular solution of the form
\begin{align}
   a =  \left[ \frac{2(m+1)}{r(m-1)^2}\right]^{\frac{1}{m -1}}  \; t^{\frac{2}{1-m}}\;. 
\end{align}
Thus, comparing Equations \eqref{SpcaseEmden} and \eqref{R2dot}, we find that
$m = 2\beta - 5$ and $r= \ga$, and $a$ reads   
\begin{align}
   a =  \left[(\beta - 3)\; \sqrt{ \frac{\ga}{\beta -2 } } \;\; t \right]^{\frac{1}{3 - \beta}}\;.\label{Rvalue} 
\end{align}

Although this is a particular solution, this still is a direct and exact generalization result of all work that has been done
on Bianchi-$V$ cosmological models with varying $\Lambda$ and $G$.
Having known the values of $a$, the dynamical and kinematical parameters of the model can be computed as follows. 
Using Equation \eqref{Rvalue}, the Hubble parameter is 
\begin{align}
H & = \frac{\dot{a}}{a} = \frac{1}{(3-\beta)}\;\frac{1}{t} ~.\label{Hubblevalue}
\end{align}
Integrating Equation \eqref{EvolEq} gives an expression for the density as  
\begin{align}
\rho & = \frac{\rho_0}{a^{6}} = \frac{\rho_0}{ \left[  (\beta - 3)\; \sqrt{ \frac{\ga}{\beta -2 } } \;\; t \right]^{\frac{6}{3 - \beta}} }~.
\label{dens}
\end{align}
Using Equation \eqref{Hubblevalue} in Equation \eqref{shear}, the expansion scalar is 
\begin{align}
\theta &= 3H =  \frac{3}{(3-\beta)}\;\frac{1}{t} ~,\label{finaltheta}
\end{align}
and, from Equations \eqref{Rvalue} and \eqref{volumeexpansion}, the shear scalar reads
\begin{align}
\sigma &= \frac{K}{a^3} = \frac{K}{\left [  (\beta - 3)\; \sqrt{ \frac{\ga}{\beta -2 } } \;\; t \right ]^{\frac{3}{3 - \beta}} }\;.
\end{align}
From Equation \eqref{Rvalue}, we can compute $\dot{a}$ and $\ddot{a}$ and substitute them into Equation \eqref{generlizedq}
to compute the constant deceleration parameter as 
\begin{align}
q &= - \frac{\ddot{a}\;a}{\dot{a^2}} = 2-\beta \;.
 \end{align}
Thus, this model can have $q>0$ (deceleration) if $\beta<2$, $q<0$ (acceleration) if $\beta>2$ and $q=0$ (constant expansion) if $\beta=2$. 

 The direct substitution of $a$ in Equation \eqref{Lambda} gives 
\begin{align}
\Lambda & = \alpha \left[  (\beta - 3)\; \sqrt{ \frac{\ga}{\beta -2 } } \;\; t \right]^{\frac{2}{\beta-3}} 
+  \frac{\beta }{(3-\beta)^2}\;\frac{1}{t^2}~,
 \label{Lfinal}
 \end{align}
 and differentiating Equation \eqref{Lfinal} with respect to $t$ gives
 \begin{align}
\dot{ \Lambda}  & = - 2\alpha \; \sqrt{ \frac{\ga}{\beta -2 }} \left[  (\beta - 3)\; \sqrt{ \frac{\ga}{\beta -2 } } \;\; t \right]^{\frac{5-\beta}{\beta-3}} 
- \frac{2\beta }{(3-\beta)^2}\;\frac{1}{t^3} \;.
 \end{align}

We use this result together with Equation \eqref{dens} to compute the Newtonian ``constant" as
 \be
G = -\int \frac{\dot{\Lambda}}{8\pi \rho}dt=G_0-\frac{\ga^2(\beta-3)^4}{8\pi\rho_0(\beta-2)^3}\left[\ga(\beta-3)t^{\frac{2\beta}{3-\beta}}-\frac{\alpha(\beta-2)}{2}t^{\frac{4}{3-\beta}}\right]\;,
\ee
and, from Equations \eqref{AEq}, \eqref{Beq} and \eqref{Ceq}, the metric variables can be computed as
\begin{align}
A&= \left[ (\beta - 3)\; \sqrt{ \frac{\ga}{\beta -2 } } \;\; t \right]^{\frac{1}{3 - \beta}} ~,\\
B&=B_0 \left[  (\beta - 3)\; \sqrt{ \frac{\ga}{\beta -2 } } \;\; t \right]^{\frac{1}{3 - \beta}}
\;\exp \left\{    \frac{k(\beta-3)}{3} \; \left[ (\beta - 3)\;\sqrt{ \frac{\ga}{\beta -2} } \; \right]^{ \frac{3}{\beta-3} } 
            t^{ \frac{ \beta}{3-\beta}} \right\} ~,\\
 C&=C_0 \left[  (\beta - 3)\; \sqrt{ \frac{\ga}{\beta -2 } } \;\; t \right]^{\frac{1}{3 - \beta}}
\;\exp \left\{     - \frac{k(\beta-3)}{3} \; \left[ (\beta - 3)\;\sqrt{ \frac{\ga}{\beta -2} } \; \right]^{ \frac{3}{\beta-3} } 
            t^{ \frac{ \beta}{3-\beta}} \right\} ~,
\end{align}
where $B_0=m_1e^{\mbox{const}}\;, C_0=m_2e^{\mbox{const}}\;, \rho_0$ and $G_0$ are constants
of integration whose actual values can be chosen using appropriate initial conditions. 

\begin{figure}[h!]
  \centering
    \subfloat[the evolution of $a(t)$ in time.]{\includegraphics[width=0.45\textwidth]{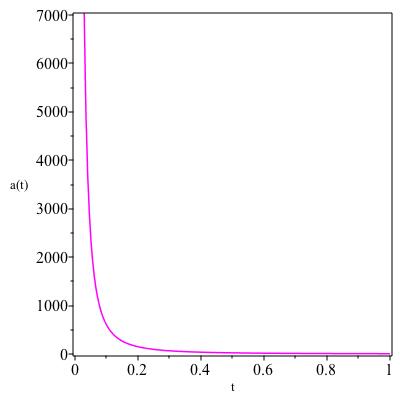}\label{figa2}}
  \subfloat[the evolution of $H(t)$ in time.]{\includegraphics[width=0.45\textwidth]{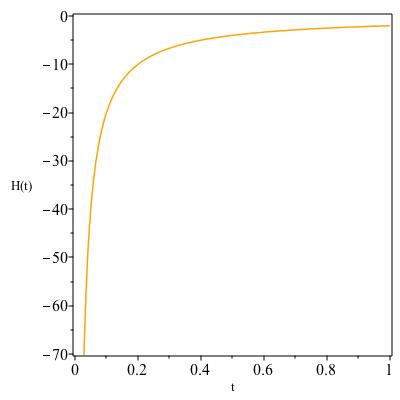}\label{figh2}}
  \caption{The variation of $a$ and $H$ for Model II. This model depicts a contracting universe (Panel (a)) for a particular set of parameters ($\alpha=\ga=1\;,\beta=3.5$, for example). Panel (b) shows that contraction occurs at a decreasing rate, asymptotically leading to a universe with vanishingly small constant size at late times.}
\label{fig8}
\end{figure}

\begin{figure}[h!]
  \centering
  \subfloat[the evolution of $\Lambda(t)$ in time.]{\includegraphics[width=0.45\textwidth]{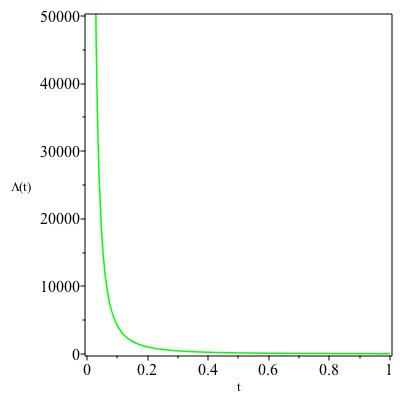}\label{figl2}}
  \subfloat[the evolution of $G(t)$ in time.]{\includegraphics[width=0.45\textwidth]{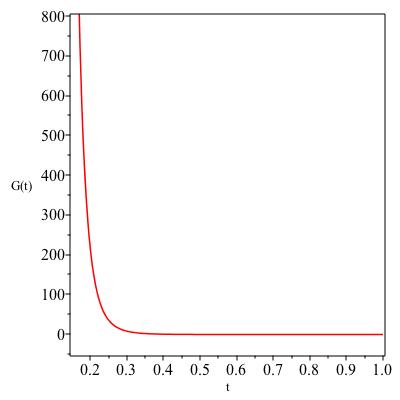}\label{figg2}}
  \caption{The variation of $\Lambda$ and $G$ for Model II. Here both $\Lambda$ and $G$ decrease in time, the former faster than the later.}
    \label{fig9}
\end{figure}

\begin{figure}[h!]
  \centering
  \subfloat[the evolution of $V(t)$ in time.]{\includegraphics[width=0.45\textwidth]{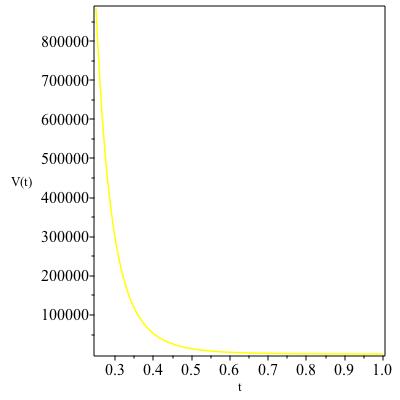}\label{figv2}}
  \subfloat[the evolution of $\sigma(t)$ in time.]{\includegraphics[width=0.45\textwidth]{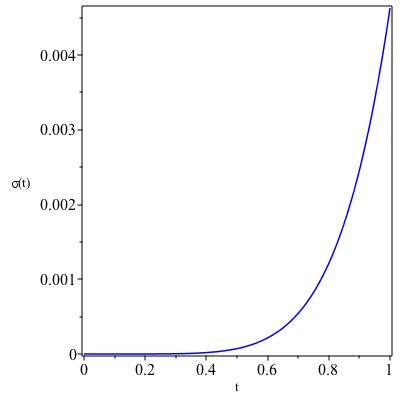}\label{figs2}}
  \caption{The variation of $V$ and $\sigma$ for Model II. Since the model has a contracting solution, we see that the overall volume of the universe decreases as expected, with increasing anisotropy at late times.}
  \label{fig10}
\end{figure}

The above results show that the model has a point of initial singularity or a Big Bang at $t=0$.
This result improves on the result obtained in \cite{Dwivedi2012}, 
where the singularity occurs at  a negative point in time. As Figures \ref{fig8}--\ref{fig10} depict, this model represents a contracting universe solution where $V$ and both $\Lambda$ and $G$ simultaneously decrease, while $H$, $\theta$, $\sigma$ and $\rho$ all increase as time increases, with $H$ asymptotically approaching zero from below. It is worth mentioning here that the model produces a constant deceleration of the expansion for acceptable values of $\beta$.
In this model, as the expansion rate $\theta$ of the universe slows down as time increases, with the expansion eventually stopping, the ratio of $|\sigma/\theta| \to \infty$ as $t\to \infty$, 
thus predicting that the universe in this model becomes anisotropic at late times. These anisotropic predictions contradict the results reported by Dwivedi and Tiwari \cite{Dwivedi2012}. Moreover, as $t\to \infty$, the metric components $A$, $B$ and $C$ approach zero rapidly, and the average scale factor $a$ tends to zero, which, with an even more rapidly increasing energy density, potentially results in a Big~Crunch.

\unskip
\unskip

\section{Conclusions}\label{conc}
The main objective of this paper was to find general exact solutions for Bianchi Type-$V$ cosmological models for a stiff perfect fluid with time-varying cosmological
and gravitational ``constants''. In practice, this is a generalization process of previous works without prior choice of  the quantum field theoretically adjustable parameters $\alpha$ and $\beta$ that define the cosmological constant as $\Lambda = \alpha/a^{2} + \beta H^2$.  

In our solution process, we have followed two steps: (i) the EFEs were derived and reduced to a coupled system of differential equations, plus arbitrary constants of
integration. This provides a way to integrate for the values of $A$, $B$, $C$,
$\rho$, $\Lambda$ and $G$, but it is not a complete solution because of the unknown value of $a$; (ii) The EFEs have been transformed to a nonlinear second-order DE in $a$, whose solution has been obtained using the hypergeometric function.
Throughout the solution steps, we did not assume a priori the values
of $\alpha$ and $\beta$, unlike previous works in the literature. We have therefore tried to solve the system in a more general setting. We have shown that the hypergeometric function is controlled by the value of $\beta$ and makes it very difficult to have a closed form for $a$ as an explicit function of $t$.

Finally, two cosmological models are obtained with a choice of suitably fixed values of $\beta$ and 
through the transformation of the generalized Friedman equation into a special case of the Emden--Fowler equation.
The dynamical and kinematical parameters of each model are exactly computed and clearly discussed. It has been shown that, while one of these models results in
a universe that asymptotically isotropizes at late times, the other becomes increasingly anisotropic. 

As more precise data become available, it will, {\it in principle}, be possible to constrain the integration constants that we chose arbitrarily in this study to get a better picture of this class of cosmological models. We leave this particular task to a future consideration.
\vspace{12pt}

\ack 
This work is based on the research supported in part by the National Research Foundation of South Africa (Grant Numbers 109257 and 112131). AA acknowledges the hospitality of the Department of Physics, University of Khartoum, during whose visit the collaboration on this project was initiated.

\end{document}